\documentclass[aps,12pt,pra,superscriptaddress,preprint,a4paper]{revtex4-1}
\usepackage{amsmath}
\usepackage{subfig}
\usepackage{array}
\usepackage{amsfonts}
\usepackage{epsf}
\usepackage{epsfig}
\usepackage{float}
\usepackage{epstopdf}
\usepackage{amssymb}
\usepackage{bbm}
\usepackage{xcolor}
\numberwithin{equation}{section}
\begin{document}
\begin{center}
%\fbox{\small Paper for the journal: J. Math. Anal. Appl}\\
\end{center}
\title{Higher dimensional fractional time independent Schr\"{o}dinger equation via Jumarie fractional derivative with generalized pseudoharmonic potential}
\author{\small Tapas Das }
\email[E-mail: ]{tapasd20@gmail.com}\affiliation{Kodalia Prasanna Banga High School (H.S), South 24 Parganas 700146, India}
\author{\small Uttam Ghosh }
\email[E-mail: ]{uttam_math@yahoo.co.in}
\author{\small Susmita Sarkar}
\email[E-mail: ]{susmita62@yahoo.co.in}\affiliation{Department of Applied Mathematics, University of Calcutta, Kolkata, India}
\author{\small Shantanu Das}
\email[E-mail: ]{shantanu@barc.gov.in}\affiliation{Reactor Control System Design Section (E \& I Group) BARC Mumbai India}

\begin{abstract}
In this paper we obtain approximate bound state solutions of $N$-dimensional time independent fractional Schr\"{o}dinger equation for generalised pseudoharmonic potential which has the form $V(r^{\alpha})=a_1r^{2\alpha}+\frac{a_2}{r^{2\alpha}}+a_3$. Here $\alpha(0<\alpha<1)$ acts like a fractional parameter for the space variable $r$. The entire study is composed with the Jumarie type derivative and the elegance of Laplace transform. As a result we successfully able to express the approximate bound state solution in terms of Mittag-Leffler function and fractionally defined confluent hypergeometric function. Our study may be treated as a generalization of all previous works carried out on this topic when $\alpha=1$ and $N$ arbitrary. We provide numerical result of energy eigenvalues and eigenfunctions for a typical diatomic molecule for different $\alpha$ close to unity. Finally, we try to correlate our work with Cornell potential model which corresponds to $\alpha=\frac{1}{2}$ with $a_3=0$ and predict the approximate mass spectra of quarkonia. 
\end{abstract}
\maketitle
\section{I\lowercase{ntroduction}}
Although the history of fractional calculus is almost as long as that of integer-order calculus, for many years they were not used in physics and applied sciences. The reason of such unpopularity could be that there exist several other definitions of fractional derivatives [1] as well as a lack of geometrical interpretation of them [2]. The situation changed after 1970 when Mandelbrot [3] proposed fractional dimension and a close inter connection between Brownian motion and Riemann-Liouville fractional calculus. From then on, fractional calculus started to attract physicists to explore the complex phenomena originated from different dissipative forces present in the nature. To that aim Riewe [4] introduced fractional Hamiltonian and Lagrangian equations of motion for non-conservative systems. This insisted physicists to develop fractional Schr\"{o}dinger equation, because Hamiltonian canonical equations are the basic starting theory of non-relativistic quantum mechanics. Using fractional canonical equation of motion Muslih \textit{et al} [5] derived fractional Schr\"{o}dinger equation containing partial left and right Riemann-Liouville fractional derivatives. Later Laskin [6-7] able to generalise the Feynman path integral to L\'{e}vy one and developed the space fractional Schr\"{o}dinger equation. Soon after Laskin, Guo and Xu [8], Dong and Xu [9] studied the space fractional Schr\"{o}dinger equation with few specific potential models.\\
Despite of all these studies still there is a dilemma to use Riemann-Liouville derivative rules in quantum mechanics as these rules are not quite parallel to the well known classical calculus [10]. The embarrassing fact of classical Riemann-Liouville derivative is that fractional derivative of a constant is not zero. Though Caputo [11] derivative solved the problems but it has own disadvantage as it cannot work on non-differentiable functions. To get rid of the above problem Jumarie [12] modified the Riemann-Liouville derivative that can run parallel with the classical calculus rules. In quantum mechanics the Jumarie type derivative rules are always welcome because quantum mechanics in general deals with predetermined boundary values of wave function or its derivative on a boundary as a zero. During last few years use of Jumarie type derivative and its application in various fields including quantum mechanics studied a lot [13-19].\\
Recently, we have elaborately studied the $N$-dimensional Schr\"{o}dinger equation for Mie-type spherical symmetric potential [27] composed with Jumarie type fractional derivative. The study was mainly devoted to realize the fractional Laplacian operator in hyperspherical coordinate system in $N$-spatial dimensions and tried to find the solution of $N$-dimensional Schr\"{o}dinger equation for bound state of generalized Mie-type potential $V(r^{\alpha})=\frac{A}{r^{2\alpha}}+\frac{B}{r^{\alpha}}+C$ characterized by a fractional parameter $\alpha$. To achieve the goal we have used the rules of Laplace transformation of fractional differ-integrals. Finally going through the rigorous mathematics of fractional differential equation with Jumarie type differential coefficient, under well behaved boundary conditions, we have succeeded to obtain approximate bound state solution of the potential for $\alpha\approx 1.00$. \\ 
Motivated by our previous work, in this paper we have just replaced the generalized Mie-type potential with generalized pseudoharmonic potential which has same popularity as Mie-type in molecular and chemical physics. The main reason behind the study of  generalized pseudoharmonic potential separately is: under substitution the potential into the Schr\"{o}dinger equation, resulting fractional differential equation does not offer an easy way to solve it. Even Laplace transform becomes quite tedious if above said fractional differential equation is not manipulated properly. Keeping all these odds, in this paper we have approached to the generalised pseudoharmonic potential which can be written as         
\begin{eqnarray}
V(r^{\alpha})=a_1r^{2\alpha}+\frac{a_2}{r^{2\alpha}}+a_3\,,
\end{eqnarray}
where $a_i(i=1,2,3)$ are some suitable constants. When $\alpha=1$ this potential converts into the original form of pseudoharmonic potential [20].
\begin{eqnarray}
V(r)=D_0\Big(\frac{r}{r_e}-\frac{r_e}{r}\Big)^2\,.
\end{eqnarray}
The symbol $D_0$ stands for the dissociation energy and it is given by $D_0=\frac{1}{8}K_er_e^2$ where $K_e$ is force constant due to the bonding of the diatomic molecule and $r_e$ is the equilibrium constant for the same. \\
The present paper is organized as follows: In the next section we shall provide a very short note on the Jumarie type fractional derivative and Laplace transform of fractional differ-integrals. Section 3 is devoted for the bound state spectrum for the pseudoharmonic potential. Discussion appears in section 4 where theoretical as well as numerical results are discussed with few eigenfunctions plotting. We have also furnished approximate mass spectra of quarkonia via the Cornell potential model, which is equivalent to our potential model corresponds to $\alpha=0.5$. Finally the conclusion of the work comes in the section 5. 
\section{O\lowercase{utline of fractional derivative and} L\lowercase{aplace transform}}
\subsection{Fractional order derivative of Jumarie type}
Jumarie [12-14] defined the fractional order derivative by modifying the left-Riemann-Liouville (RL) fractional derivative in the following form for a continuous function $f(x)$ (but not necessarily differentiable) in the interval $a$ to $x$, with $f(x)=0$ for $x<a$
\begin{eqnarray}
\,_a^{J}D_{x}^{\alpha}[f(x)]=f^{(\alpha)}(x)=\begin{cases} \frac{1}{\Gamma(-\alpha)}\int_a^{x}(x-\xi)^{-\alpha-1}f(\xi)d\xi,&   \alpha<0,\\
\frac{1}{\Gamma(1-\alpha)}\frac{d}{dx}\int_a^{x}(x-\xi)^{-\alpha}(f(\xi)-f(a))d\xi,&  0<\alpha<1,\\
(f^{(\alpha-n)}(x))^{(n)},&  n \leq\alpha<n+1.
\end{cases}
\end{eqnarray}
It is customary to take the start point of the interval as $a=0$ and use the symbol $\,_0^{J}D_{x}^{\alpha}[f(x)]$ for $f^{(\alpha)}(x)$. Here from in the rest of the paper we will always denote the fractional derivative $f^{(\alpha)}(x)\equiv\frac{d^\alpha f(x)}{dx^\alpha}$ as $\,_0^{J}D_{x}^{\alpha}[f(x)]$ with Jumarie sense. In the above definition, the first expression is just the Riemann-Liouville fractional integration, the second expression is known as modified Riemann-Liouville derivative of order $0<\alpha<1$ because of the involvement of $f(a)$. The third line definition is for the range $n\leq\alpha<n+1$. Apart from the integral type of definition we can also express fractional derivative via fractional difference. Let $f:\Re\rightarrow\Re$, denotes a continuous (but not necessarily differentiable) function such that $x\rightarrow f(x)$ for all $x\in\Re$. If $h>0$ denotes a constant discretization span with forward operator $FW(h)f(x)=f(x+h)$; then the right hand fractional difference of $f(x)$ of order $\alpha$ ($0<\alpha<1$) is defined by the expression [21]
\begin{eqnarray}
\bigtriangleup^{\alpha}f(x)=(FW(h)-1)^{\alpha}f(x)=\sum_{i=0}^{\infty}(-1)^i{{\alpha}\choose{i}}f[x+(\alpha-i)h]\,,
\end{eqnarray}
where generalized binomial coefficients $\frac{\Gamma(\alpha-i)}{\Gamma(-\alpha)\Gamma(i+1)}={{i-\alpha-1}\choose{i}}=(-1)^i{{\alpha}\choose{i}}$. These equalities being readily established from the definition of a binomial coefficient and generalization of factorials with Gamma function $\,^nC_r={{n}\choose{r}}=\frac{n!}{r!(n-r)!}$. 
Then the Jumarie fractional derivative is defined as
\begin{eqnarray}
f^{(\alpha)}_+(x)=\lim_{h\downarrow 0}\frac{\bigtriangleup^{\alpha}_+[f(x)-f(0)]}{h^{\alpha}}=\frac{d^{\alpha}f(x)}{dx^{\alpha}}\,.
\end{eqnarray}
This definition is close to the standard definition of derivatives for beginner's study. Following this definition it is clear that the $\alpha$-th derivative of a constant for $0<\alpha<1$ is zero. Few results for Jumarie type derivative are listed below depending on the characteristics of given function $(f[u(x)])$ [22]
\begin{subequations}
\begin{align}
\,_0^{J}D_{x}^{\alpha}(f[u(x)])&=f_{u}^{(\alpha)}(u)(u_x^{'})^{\alpha}\,,\\
\,_0^{J}D_{x}^{\alpha}(f[u(x)])&=(f/u)^{1-\alpha}(f^{'}_u(u))^{\alpha}u^{\alpha}(x)\,,\\
\,_0^{J}D_{x}^{\alpha}(f[u(x)])&=(1-\alpha)!u^{\alpha-1}f_u^{(\alpha)}(u)u^{\alpha}(x)\,,\\
\,_0^{J}D_{x}^{\alpha}(x^{\beta})&=\frac{\Gamma(1+\beta)}{\Gamma(1+\beta-\alpha)}x^{\beta-\alpha}\,. 
\end{align}
\end{subequations}
In fractional calculus solution of any linear fractional differential equation, composed with Jumarie derivative, can be easily obtained in terms of Mittag-Leffler function of one parameter [23] which is defined as
\begin{eqnarray}
E_{\alpha}(z)=\sum_{\kappa=0}^{\infty}\frac{z^\kappa}{\Gamma(\alpha \kappa+1)}\,,\,\, (\alpha>0)\,.
\end{eqnarray} 
or more general form [24] $E_{\alpha,\beta}(z)=\sum_{\kappa=0}^{\infty}\frac{z^\kappa}{\Gamma(\alpha \kappa+\beta)}$. Clearly $E_{\alpha,1}(z)=E_{\alpha}(z)$ and $E_{1,1}(z)=E_1(z)=e^z$. We provide few derivative rules [17-18] associated with the Mittag-Leffler function and its trigonometric counterparts.
\begin{subequations}
\begin{align}
\,_0^{J}D_{x}^{\alpha}[E_{\alpha}(ax^{\alpha})]&=aE_{\alpha}(ax^{\alpha})\,,\\
\,_0^{J}D_{x}^{\beta}[E_{\alpha}(ax^{\alpha})]&=x^{\alpha-\beta}E_{\alpha,\alpha-\beta+1}(x^{\alpha})\,,\\
\,_0^{J}D_{x}^{\alpha}[cos_{\alpha}(ax^{\alpha})]&=-asin_{\alpha}(ax^{\alpha})\,,\\
\,_0^{J}D_{x}^{\alpha}[sin_{\alpha}(ax^{\alpha})]&=acos_{\alpha}(ax^{\alpha})\,,
\end{align}
\end{subequations}
where one parameter fractional sine and cosine function are defined as follows [14]\\
$cos_{\alpha}(x^{\alpha})=\sum_{\kappa=0}^{\infty}(-1)^\kappa \frac{x^{2\kappa\alpha}}{\Gamma(1+2\alpha \kappa)}$ and $sin_{\alpha}(x^{\alpha})=\sum_{\kappa=0}^{\infty}(-1)^\kappa \frac{x^{(2\kappa+1)\alpha}}{\Gamma(1+(2\kappa+1)\alpha )}$ with \\$E_{\alpha}(ix^{\alpha})=cos_{\alpha}(x^{\alpha})+isin_{\alpha}(x^{\alpha})$.  
\subsection{Laplace transformation of fractional differ-integrals}
In general Laplace transform $F(s)$ or $\mathcal{L}$ of a function $f(x)$ is defined as [25]
\begin{eqnarray}
F(s)=\mathcal{L}\left\{f(x)\right\}=\int_0^\infty e^{-sx}f(x)dx\,.
\end{eqnarray}
If there is some constant $\sigma\in\Re$ such that $|e^{-\sigma x}f(x)|\leq\mathcal{M}$ for sufficiently large $x$, the above definition will exist for Re $[s]>\sigma$. The following are the well known derivative properties of Laplace transform when $n$ is an integer.
\begin{subequations}
\begin{align}
\mathcal{L}\left\{f^{(n)}(x)\right\}&=s^nF(s)-\sum_{k=0}^{n-1}s^{n-1-k}{f^{(k)}(0)}\,,\\
\mathcal{L}\left\{x^{n}f(x)\right\}&=(-1)^{n}F^{(n)}(s)\,,
\end{align}
\end{subequations}
where the superscript $(n)$ denotes the $n$-th derivative with respect to $x$ for $f^{(n)}{(x)}$, and with respect to $s$ for $F^{(n)}(s)$. Now if $n$ becomes non integer, say $\alpha$, then above two rules are generalized as [26]
\begin{subequations}
\begin{align}
\mathcal{L}\left\{\frac{d^{\alpha}f(x)}{dx^{\alpha}}\right\}&=s^\alpha F(s)-\sum_{k=0}^{n-1}s^k\frac{d^{\alpha-1-k}f(x)}{dx^{\alpha-1-k}}\Bigg|_{x=0}\,,\\
\mathcal{L}\left\{x^{\alpha}f(x)\right\}&=-\tau\frac{d^\alpha F(s)}{ds^\alpha}\,,
\end{align}
\end{subequations}
where, $n$ is the largest integer such that $(n-1)<\alpha\leq n$ and $\tau=-\frac{cosec((\alpha-\delta)\pi)}{cosec(-\delta \pi)}$ with $-1<\delta<0$ [27]. The sum in the Eq.(2.9a) is zero when $\alpha\leq 0$.  The Eq.(2.9b) shows that $\mathcal{L}\left\{x^{\alpha}f(x)\right\}\neq \pm\frac{d^\alpha F(s)}{ds^\alpha}$ but when $\alpha=1$, (that makes $\tau=1$) this is consistent with Eq.(2.8b) if we take $n=1$. Choosing the initial condition $f(0)=0$ (frequently appears in quantum mechanical problems) it is easy to have $\mathcal{L}\left\{\frac{d^{\alpha}f(x)}{dx^{\alpha}}\right\}=s^\alpha F(s)$. Under this circumstance one can generate $\mathcal{L}\left\{x^\alpha\frac{d^{\beta}f(x)}{dx^{\beta}}\right\}$ as follows
\begin{align}
\mathcal{L}\left\{x^\alpha\frac{d^{\beta}f(x)}{dx^{\beta}}\right\}&=-\tau\frac{d^\alpha}{ds^\alpha}\mathcal{L}\left\{\frac{d^{\beta}f(x)}{dx^{\beta}}\right\}\,,\nonumber\\
&=-\tau\frac{d^\alpha}{ds^\alpha}[s^\beta F(s)]\nonumber \\
&=-\tau\Big[F(s)\frac{\Gamma(1+\beta)}{\Gamma(1+\beta-\alpha)}s^{\beta-\alpha}+s^\beta\frac{d^\alpha F(s)}{ds^\alpha}\Big]\,,
\end{align}
where we have used the rule $(uv)^\alpha=u^{(\alpha)}v+v^{(\alpha)}u$ in Jumarie sense. This operational formula will be used in the next section. It is necessary to mention here that the condition $f(0)=0$ not only goes with Jumarie sense but also with the other fractional derivative formulation, like Caputo. These two fractional derivatives are same if and only if the function being considered is differentiable. Fractional derivative of Jumarie type does not demand the function need to be differentiable, whereas the Caputo definition demand the condition of differentiability.  There are few more operational formulas which are well established in the literature [1,26]
\begin{subequations}
\begin{align}
\mathcal{L}\left\{x^\alpha\right\}&=\frac{\Gamma(1+\alpha)}{s^\alpha}\,,\\
\mathcal{L}\left\{E_\alpha(ax^\alpha)\right\}&=\frac{s^{\alpha-1}}{s^\alpha-a}\,,\\
\mathcal{L}\left\{E_\alpha^{(k)}(-x^\alpha)\right\}&=\frac{s^{\alpha+k-1}}{s^\alpha+1}\,,\\
\mathcal{L}\left\{x^{\alpha k+\beta-1}E_{\alpha,\beta}^{(k)}(\pm ax^\alpha)\right\}&=\frac{k!s^{\alpha-\beta}}{(s^{\alpha}\mp a)^{k+1}}\,,
\end{align}
\end{subequations}
\section{B\lowercase{ound state spectrum of fractional} P\lowercase{seudoharmonic potential}}
If we choose the natural unit $\hbar=c=1$ then for large-$N$ expansion [33], $N$-dimensional fractional time independent Schr\"{o}dinger equation for a diatomic molecule in centre of mass coordinate is written as [27],
\begin{eqnarray}
\Big[\nabla_{N}^{2\alpha}+2M(\mathcal{E}_{\alpha}-V(r^{\alpha}))\Big]\psi(r^{\alpha},\Omega^{\alpha}_N)=0\,,
\end{eqnarray} 
$M=\frac{m_1m_2}{m_1+m_2}$ denotes the reduced mass of the molecule where $m_1$ and $m_2$ are the masses of constituent particles forming the molecule. Furthermore $\mathcal{E}_{\alpha}$ and $V(r^{\alpha})$ are the fractional energy eigenvalue and fractional potential energy respectively. They both have unit $GeV^{\alpha}$. We will consider the mass $M$ as a fractional mass also with energy unit $GeV^{\alpha}$. When $\alpha=1$, all these units are well familiar within the natural unit scheme. The term $\Omega^{\alpha}_N$ within the argument of $\psi$ denotes angular variables $\theta_1^{\alpha}, \theta_2^{\alpha}, \theta_3^{\alpha} \cdots \theta_{N-2}^{\alpha}, \phi^{\alpha}$ [27]. The term 
$\nabla_{N}^{2\alpha}$ is called fractional Laplacian operator in $N$ dimension. In terms of hyperspherical coordinates it can be further written as 
\begin{equation}
\nabla_{N}^{2\alpha}=\frac{1}{(\alpha!)^2}\frac{1}{(r^{\alpha})^{N-1}}\frac{\partial^{\alpha}}{\partial r^{\alpha}}\Big[(r^{\alpha})^{N-1}\frac{\partial^{\alpha}}{\partial r^{\alpha}}\Big]-\frac{\Lambda^{2\alpha}_{N-1}}{r^{2\alpha}}\,,
\end{equation} 
where $\Lambda^{2\alpha}_{N-1}$ is fractional hyperangular momentum operator. The explicit form is
\begin{eqnarray}
\Lambda^{2\alpha}_{N-1}=-\Bigg[\sum_{k=1}^{N-2}\frac{1}{sin^{2}_{\alpha}\theta^{\alpha}_{k+1}sin^{2}_{\alpha}\theta^{\alpha}_{k+2}\cdots sin^{2}_{\alpha}\phi^{\alpha}}\Big(\frac{1}{sin^{k-1}_{\alpha}\theta^{\alpha}_{k}}\frac{\partial^{\alpha}}{\partial\theta_k^{\alpha}}sin^{k-1}_{\alpha}\theta^{\alpha}_{k}\frac{\partial^{\alpha}}{\partial\theta_k^{\alpha}}\Big)+\nonumber\\ \frac{1}{sin^{N-2}_{\alpha}\phi^{\alpha}}\frac{\partial^{\alpha}}{\partial\phi^{\alpha}}\Big(sin^{N-2}_{\alpha}\phi^{\alpha}\frac{\partial^{\alpha}}{\partial\phi^{\alpha}}\Big)\Bigg]\,.
\end{eqnarray} 
Taking the solution by means of separation variable technique 
$\psi(r^{\alpha},\Omega^{\alpha}_N)=R(r^{\alpha})Y(\Omega^{\alpha}_N)$ and adopting the eigenvalue equation for $Y(\Omega^{\alpha}_N)$ as 
\begin{eqnarray}
\Lambda^{2\alpha}_{N-1}Y(\Omega^{\alpha}_N)=\ell(\ell+N-2)|_{N>1}Y(\Omega^{\alpha}_N)\,,
\end{eqnarray}
where $\ell$ is orbital angular momentum quantum number (can take quantized values $0,1,2,3\cdots$ only), we have the fractional order hyperradial or in short `radial' equation [27]
\begin{eqnarray}
\Bigg[\frac{d^{2\alpha}}{dr^{2\alpha}}+\frac{\Gamma(1+\alpha(N-1))}{\Gamma(1+\alpha(N-2))}\frac{1}{r^{\alpha}}\frac{d^{\alpha}}{dr^{\alpha}}-\frac{\ell(\ell+N-2)(\alpha!)^2}{r^{2\alpha}}\nonumber\\+2M(\alpha!)^2(\mathcal{E}_{\alpha}-V(r^{\alpha}))\Bigg]R(r^{\alpha})=0\,.
\end{eqnarray} 
Inserting the potential (1.1) into Eq.(3.5) we have
\begin{eqnarray}
\,_0^{J}D_{r}^{2\alpha}[R(r^{\alpha})]+\frac{\Gamma(1+\alpha(N-1))}{\Gamma(1+\alpha(N-2))}\frac{1}{r^{\alpha}}\,_0^{J}D_{r}^{\alpha}[R(r^{\alpha})]\nonumber\\+\Big[\epsilon_\alpha^2-\mu_\alpha^2 r^{2\alpha}-\frac{\nu_\alpha(\nu_\alpha+1)}{r^{2\alpha}}\Big]R(r^{\alpha})=0\,,
\end{eqnarray}
where
\begin{subequations}
\begin{align}
\nu_{\alpha}(\nu_{\alpha}+1)&=\ell(\ell+N-2)(\alpha!)^2+2M(\alpha!)^2 a_2\,,\\
\epsilon_{\alpha}^2&=2M(\alpha!)^2(\mathcal{E}_{\alpha}-a_3)\,,\\
\mu_{\alpha}^2&=2M(\alpha!)^2 a_1\,,
\end{align}
\end{subequations}
Quantum mechanical bound state eigenfunctions are generally well behaved in nature, that means they are bounded or $\psi(r^{\alpha},\Omega^{\alpha}_N)$ approach to zero for $r\rightarrow 0$ and $r\rightarrow \infty$. Since this type of initial conditions are associated with $R(r^{\alpha})$, we predict the solution of Eq.(3.6) as 
\begin{eqnarray}
R(r^{\alpha})=(r^{\alpha})^{-k}f(r^{\alpha})|_{k>0}\,.
\end{eqnarray}  
Here the term $(r^{\alpha})^{-k}$ ensures the fact that $R(r\rightarrow \infty)=0$. The unknown function $f(r^{\alpha})$ is expected to behave like $f(r\rightarrow 0)=0$. After deriving $\,_0^{J}D_{r}^{2\alpha}[f(r^{\alpha})]$, $\,_0^{J}D_{r}^{\alpha}[f(r^{\alpha})]$ and performing little calculation on Eq.(3.6) we have
\begin{eqnarray}
\,_0^{J}D_{r}^{2\alpha}f(r^{\alpha})+\frac{Q_{1}(\alpha,k,N)}{r^{\alpha}}\,_0^{J}D_{r}^{\alpha}f(r^{\alpha})+\Big[\frac{Q_{2}(\alpha,k,N,\nu_{\alpha})}{r^{2\alpha}}-\mu_\alpha^2r^{2\alpha}+\epsilon_{\alpha}^2\Big]f(r^{\alpha})=0\,,
\end{eqnarray}
where
\begin{subequations}
\begin{align}
 Q_{1}(\alpha,k,N)&=\frac{2\Gamma(1-\alpha k)}{\Gamma(1-\alpha k-\alpha)}+\frac{\Gamma(1+\alpha(N-1))}{\Gamma(1+\alpha(N-2))}\,,\\
 Q_{2}(\alpha,k,N,\nu_{\alpha})&=\frac{\Gamma(1-\alpha k)}{\Gamma(1-\alpha k-\alpha)}\Bigg[\frac{\Gamma(1-\alpha k-\alpha)}{\Gamma(1-\alpha k-2\alpha)}+\frac{\Gamma(1+\alpha(N-1))}{\Gamma(1+\alpha(N-2))}\Bigg]-\nu_{\alpha}(\nu_{\alpha}+1)\,.
\end{align}
\end{subequations}
Finding the solution of Eq.(3.9) is a difficult task due to the strong singular term $\frac{Q_{2}(\alpha,k,N,\nu_{\alpha})}{r^{2\alpha}}$. To ease out the situation we will study the Eq.(3.9) in transformed space (Laplace) with a parametric restriction 
\begin{eqnarray}
Q_{2}(\alpha,k,N,\nu_{\alpha})=0\,,
\end{eqnarray}
without loss of any generality. It is worth to mention here that, above restriction is not a mandatory or essential to apply the Laplace transform on Eq.(3.9). It helps to avoid the tenacious mathematical steps only. Denoting the solution of above equation for $k(>0)$ as $k_{\alpha}^{*}$ we can rewrite Eq.(3.9) as
\begin{eqnarray}
\,_0^{J}D_{r}^{2\alpha}g(r)+\frac{Q_{1}(\alpha,k_{\alpha}^{*},N)}{r^{\alpha}}\,_0^{J}D_{r}^{\alpha}g(r)+(\epsilon_{\alpha}^2-\mu_\alpha^2r^{2\alpha})g(r)=0\,,
\end{eqnarray} 
where $f(r^{\alpha})$ is replaced with $g(r)$. In spite of the condition $Q_2=0$ again the present equation is not suitable for Laplace transform because, the term containing $r^{2\alpha}$ will generate higher order fractional differential equation in transformed space, which will be difficult to tackle. There is an alternative way. If we adopt a change in the variable then the situation becomes much more easy. Taking $y=r^2$ and using the rule (2.4a) as $\,_0^{J}D_{r}^{\alpha}y=\,_0^{J}D_{y}^{\alpha}y (\frac{dy}{dr})^\alpha$ we have the following operators 
\begin{subequations}
\begin{align}
\,_0^{J}D_{r}^{\alpha}&=2^{\alpha}y^{\alpha/2}\,_0^{J}D_{y}^{\alpha}\,,\\
\,_0^{J}D_{r}^{2\alpha}&=4^{\alpha}y^{\alpha}\,_0^{J}D_{y}^{2\alpha}+4^{\alpha}\frac{\Gamma(1+\frac{\alpha}{2})}{\Gamma(1-\frac{\alpha}{2})}\,_0^{J}D_{y}^{\alpha}\,.
\end{align}
\end{subequations}
These operators help to rewrite the Eq.(3.12) in a new form with $g(r)\Leftrightarrow \chi(y)$ as
\begin{eqnarray}
y^{\alpha}\,_0^{J}D_{y}^{2\alpha}\chi(y)+\Big(\frac{\Gamma(1+\frac{\alpha}{2})}{\Gamma(1-\frac{\alpha}{2})}+\frac{Q_1}{2^\alpha}\Big)\,_0^{J}D_{y}^{\alpha}\chi(y)+\frac{1}{4^\alpha}(\epsilon_\alpha^2-\mu_\alpha^2 y^\alpha)\chi(y)=0\,.
\end{eqnarray}
Now defining $\mathcal{L}\left\{\chi(y)\right\}=\varphi(s)$ and using the rules of Laplace transform, mentioned in subsection (2.2) with $\chi(0)=0$, it is easy to obtain the following fractional differential equation
\begin{eqnarray}
\,_0^{J}D_{s}^{\alpha}\varphi(s)+\eta(s^{\alpha})\varphi(s)=0\,,
\end{eqnarray}   
where 
\begin{subequations}
\begin{align}
\eta{(s^{\alpha})}&=\frac{\lambda_1}{s^{\alpha}+\frac{\mu_\alpha}{2^\alpha}}+\frac{\lambda_2}{s^{\alpha}-\frac{\mu_\alpha}{2^\alpha}}\,,\\
\lambda_1&=\frac{\gamma_\alpha}{2}+\frac{\epsilon_\alpha^2}{2^{\alpha+1}\tau^2\mu_\alpha}\,,\\
\lambda_2&=\frac{\gamma_\alpha}{2}-\frac{\epsilon_\alpha^2}{2^{\alpha+1}\tau^2\mu_\alpha}\,,\,,\\
\gamma_\alpha &=\frac{\Gamma(1+2\alpha)}{\Gamma(1+\alpha)}-\frac{1}{\tau^2}\Big(\frac{\Gamma(1+\frac{\alpha}{2})}{\Gamma(1-\frac{\alpha}{2})}+\frac{Q_1}{2^\alpha}\Big)\,.
\end{align}
\end{subequations}
The exact solution of Eq.(3.15) is very complicated in fractional domain. We have approximately solved this type of equation in our previous work [27] for $\alpha\approx 1.00$ with the help of `$\alpha$-logarithmic' function [31] in
Jumarie sense i.e, $\int\frac{d^\alpha t}{t}=Ln_\alpha(\frac{t}{C}), t=E_\alpha(Ln_\alpha t)$ where $C$ denotes a constant such that $(\frac{t}{C})>0$. The approximate solution in transformed space yields
\begin{eqnarray}
\varphi(s)=C_{1}\Big(s^{\alpha}+\frac{\mu_\alpha}{2^\alpha}\Big)^{-\lambda_1}\Big(s^{\alpha}-\frac{\mu_\alpha}{2^\alpha}\Big)^{-\lambda_2}\,,
\end{eqnarray}
where $C_1$ is the integration constant. The second factor of Eq.(3.17) is a multivalued function when the power $-\lambda_2$ is a non integer. The quantum mechanical eigenfunction must be single valued in nature. So we must take
\begin{eqnarray}
-\lambda_2=\frac{\epsilon_\alpha^2}{2^{\alpha+1}\tau^2\mu_\alpha}-\frac{\gamma_\alpha}{2}=n \,, n=0,1,2,3, \ldots
\end{eqnarray}   
The inverse transform of Eq.(3.17) will provide the solution of the problem in actual space. To that aim, we expand Eq.(3.17) with help of Eq.(3.18) as  
\begin{align}
\varphi(s)&=C_1(s^{\alpha}+\frac{\mu_\alpha}{2^\alpha})^{-n-\gamma_\alpha}(s^{\alpha}-\frac{\mu_\alpha}{2^\alpha})^{n}\,, \nonumber \\ 
&=C_1(s^{\alpha}+\frac{\mu_\alpha}{2^\alpha})^{-\gamma_\alpha}\Bigg(\frac{s^{\alpha}-\frac{\mu_\alpha}{2^\alpha}}{{s^{\alpha}+\frac{\mu_\alpha}{2^\alpha}}}\Bigg)^{n}\,, \nonumber \\ 
&=C_1(s^{\alpha}+\frac{\mu_\alpha}{2^\alpha})^{-\gamma_\alpha}\Bigg[1-\frac{\mu_\alpha}{s^\alpha+\frac{\mu_\alpha}{2^\alpha}}\Bigg]^n\,, \nonumber \\
&=C_1\sum_{j=0}^{n}\frac{n!}{j!(n-j)!}(-1)^{j}(\mu_\alpha)^{j}\Big(s^{\alpha}+\frac{\mu_\alpha}{2^\alpha}\Big)^{-(\gamma_{\alpha}+j)}\,,\nonumber \\
&=C_1\sum_{j=0}^{n}\frac{n!}{j!(n-j)!}(-1)^{j}(\mu_\alpha)^{j}\frac{1}{(s^{\alpha}+\frac{\mu_\alpha}{2^\alpha})^{m_j+1}}\,, \quad\text{where $m_j=(\gamma_{\alpha}+j-1)$}\,.
\end{align}
Using the formula given by Eq.(2.11d) for $\alpha=\beta$ we can find the inverse of Eq.(3.17) quite easily. 
\begin{align}
\chi(y)&=C_1\sum_{j=0}^{n}\frac{n!}{j!(n-j)!}(-1)^{j}(\mu_\alpha)^{j} \frac{1}{\Gamma(\gamma_{\alpha}+j)}y^{\alpha(\gamma_{\alpha}+j)-1}E_{\alpha}^{(m_j)}(-\frac{\mu_\alpha}{2^\alpha}y^{\alpha})\,, \nonumber\\
&=\frac{C_1}{\Gamma(\gamma_{\alpha})}y^{\alpha\gamma_{\alpha}-1}\sum_{j=0}^{n}\frac{n!}{j!(n-j)!}(-1)^{j} \frac{\Gamma(\gamma_{\alpha})}{\Gamma(\gamma_{\alpha}+j)}(\mu_\alpha y^\alpha)^{j}E_{\alpha}^{(m_j)}(-\frac{\mu_\alpha}{2^\alpha}y^{\alpha})\,,\nonumber\\
 &=\mathcal{N}_cy^{\alpha\gamma_{\alpha}-1}E_{\alpha}^{(m_j)}(-\frac{\mu_\alpha}{2^\alpha}y^{\alpha})\,_{1}F_{1}(-n,\gamma_{\alpha}, \mu_\alpha y^\alpha)\,,
\end{align} 
where $E_{\alpha}^{(m_j)}(-\frac{\mu_\alpha}{2^\alpha}y^{\alpha})=\frac{d^{m_j}}{dy^{m_j}}E_{\alpha}(-\frac{\mu_\alpha}{2^\alpha}y^{\alpha})=\sum_{p=0}^{\infty}\frac{(p+m_j)!}{p!}\frac{(-\frac{\mu_\alpha}{2^\alpha}y^{\alpha})^{p}}{\Gamma(\alpha p+\alpha m_j+\alpha)}$[1]. \\ 
Hence
\begin{eqnarray}
g(r)=f(r^\alpha)=\mathcal{N}_cr^{2\alpha\gamma_{\alpha}-2}E_{\alpha}^{(m_j)}(-\frac{\mu_\alpha}{2^\alpha}r^{2\alpha})\,_{1}F_{1}(-n,\gamma_{\alpha}, \mu_\alpha r^{2\alpha})\,.
\end{eqnarray}
This yields complete radial eigenfunction 
\begin{eqnarray}
R_{n\alpha N \ell}(r^{\alpha})=r^{-\alpha k_{\alpha}^{*}}f(r^\alpha)=\mathcal{N}_c r^{[\alpha(2\gamma_\alpha-k_\alpha^*)-2]}E_{\alpha}^{(m_j)}(-\frac{\mu_\alpha}{2^\alpha}r^{2\alpha})\,_{1}F_{1}(-n,\gamma_{\alpha}, \mu_\alpha r^{2\alpha})\,.
\end{eqnarray}  
where $\mathcal{N}_c=\frac{C_1}{\Gamma(\gamma_{\alpha})}$ acts like a normalization constant and $\,_{1}F_{1}$ is fractionally defined confluent hypergeometric function i.e $\,_{1}F_{1}(-n,\gamma_{\alpha}, \mu_\alpha y^{\alpha})=\sum_{j=0}^{n}\frac{n!}{j!(n-j)!}(-1)^{j} \frac{\Gamma(\gamma_{\alpha})}{\Gamma(\gamma_{\alpha}+j)}(\mu_\alpha y^\alpha)^{j}$. The energy eigenvalue equation of the potential model comes out from Eq.(3.18) as
\begin{eqnarray}
\mathcal{E}_{n\alpha N \ell}=a_3+\frac{\tau^2}{\alpha!}\Big(n+\frac{\gamma_\alpha}{2}\Big)\sqrt{\frac{2^{2\alpha+1}a_1}{M}}\,.
\end{eqnarray}
The complete eigenfunction of the $N$ dimensional fractional Schr\"{o}dinger equation for pseudoharmonic potential can be given by
\begin{eqnarray*}
\psi(r^{\alpha},\Omega^{\alpha}_N)=\sum_{n\ell m}(\mathcal{N}_c)_{n\alpha N \ell}R_{n\alpha N \ell}(r^{\alpha})Y_{\ell}^{m}(\theta_1^{\alpha}, \theta_2^{\alpha}, \theta_3^{\alpha} \cdots \theta_{N-2}^{\alpha}, \phi^{\alpha})\,,
\end{eqnarray*} 
where $\sum_{n\ell m}$ helps to express the overall solution in terms of all possible solutions (linear combination) and $m$ controls the orientation of $\ell$ with its specific values. 
\section{R\lowercase{esults and} D\lowercase{iscussion}}
In this section, at first we have shown theoretically that the results obtained in section 3 are compatible with the several special cases both for lower and higher dimension, specially when $\alpha=1$. Secondly, we have furnished numerical results of our work for a specified potential parameters in different dimensions.  At the last we will try to review the famous Cornell potential for $\alpha=0.5$.
\subsection{Isotropic harmonic oscillator potential}
In this case $a_2=a_3=0$ and $a_1=\frac{1}{2}M\omega^2$ where $\omega$ is the circular frequency of the oscillator. Hence
Eq.(3.11) provides $k(k+1)-k(N-1)-\ell(\ell+N-2)=0$ which yields $k_1^*=k_{\ell N}=\ell+N-2$ as $k>0$. So we have 
\begin{subequations}
\begin{align}
Q_1&=-2k_{\ell N}+N-1\,,\\
\gamma_1 &=2+k_{\ell N}-\frac{N}{2}\,.
\end{align}
\end{subequations}
The energy eigenvalues of the oscillator become
\begin{eqnarray}
\mathcal{E}_{n1N \ell}=(2n+\ell+\frac{N}{2})\omega\,,
\end{eqnarray}
To find the eigenfunctions we take Eq.(3.22) with $\alpha=1$
\begin{eqnarray}
\chi(y)=\mathcal{N}^{'}y^{\gamma_1-1}e^{-\frac{\mu_1}{2}y}\,_{1}F_{1}(-n,\gamma_1, \mu_1 y)\,,
\end{eqnarray}  
where $\mathcal{N}^{'}=\mathcal{N}_c(-\frac{\mu_1}{2})^{m_j}$.
Hence
\begin{eqnarray}
g(r)=\chi(y)|_{y=r^2}=\mathcal{N}^{'}r^{2\gamma_1-2}e^{-\frac{\mu_1}{2}r^2}\,_{1}F_{1}(-n,\gamma_1, \mu_1 r^2)\,.
\end{eqnarray} \\
The radial eigenfunctions in this case is
\begin{eqnarray}
R_{n1\ell N}(r)=r^{-k_{\ell N}}g(r)=\mathcal{N}^{'}r^{\ell}e^{-\frac{\mu_1}{2}r^2}\,_{1}F_{1}(-n,\ell+\frac{N}{2}, \mu_1 r^2)\,,
\end{eqnarray}
where we have used Eq.(4.1b) with $k_{\ell N}=\ell+N-2$. These all results have been already achieved in ref.[32]. Inserting $N=3$ one can get the required results for isotropic harmonic oscillator in ordinary space.
\subsection{Pseudoharmonic potential}
In this case $a_i(i=1,2,3)\neq 0$. So Eq.(3.11)provides
\begin{eqnarray}
k_{\ell N}=\frac{N}{2}-1+\frac{1}{2}\sqrt{(N+2\ell-2)^2+8Ma_2}\,.
\end{eqnarray} 
The expressions of $Q_1$, $\gamma_1$ are same as Eq.(4.1a) and Eq.(4.1b). The energy eigenvalues for pseudoharmonic potential are
\begin{eqnarray}
\mathcal{E}_{n1N \ell}=a_3+\sqrt{\frac{8a_2}{M}}\Bigg[n+\frac{1}{2}+\frac{1}{4}\sqrt{(N+2\ell-2)^2+8Ma_2}\Bigg]\,.
\end{eqnarray}
To extract the eigenfunctions we again use the Eq.(3.22) with $\alpha=1$ as previous. The expression of $g(r)$ is the same what we have derived in the Eq.(4.4) except $\gamma_1=2+k_{\ell N}-\frac{N}{2}$ where $k_{\ell N}$ is given by Eq.(4.6). Finally the radial eigenfunctions are
\begin{eqnarray}
R_{n1\ell N}(r)=r^{-k_{\ell N}}g(r)=\mathcal{N}^{'}r^{2+k_{\ell N}-N}e^{-\frac{\mu_1}{2}r^2}\,_{1}F_{1}\Big(-n,2+k_{\ell N}-\frac{N}{2},\mu_1 r^2\Big)\,,
\end{eqnarray} 
where $\mathcal{N}^{'}=\mathcal{N}_c(-\frac{\mu_1}{2})^{m_j}$ acts like a normalization constant. This all results are matched with the work cited in reference [20]. \\
Apart from verifying the earlier works for $\alpha=1.00$, we also provide numerical result of our entire model. The potential parameters $a_{i=1,2,3}$ are assigned different values close to the relevant experimental situation. Assuming the diatomic molecular mass $M=1 GeV^{\alpha}$, we have taken $a_1=10^{-3} GeV^{3\alpha}$, $a_2=0.1 GeV^{-\alpha}$ and $a_3=0$ for constructing the TABLE 2. 
\newpage
\begin{table}[http]
\begin{center}
{\bf Numerical results of energy eigenvalues of fractional pseudoharmonic potential}\\ 
\caption{$\tau$ against $\alpha$ when $\delta=-0.5$}
%\vspace{2 mm}
\renewcommand{\arraystretch}{0.5}
\begin{tabular}{|>{\centering\arraybackslash}m{1in}|>{\centering\arraybackslash}m{1in}|}\hline
$\alpha$ & $\tau$ \\ \hline
$0.70$ & $1.7013$\\ \hline
$0.75$ & $1.4142$\\ \hline
$0.80$ & $1.2361$\\ \hline
$0.85$ & $1.2223$\\ \hline
$0.90$ & $1.0515$\\ \hline
$0.95$ & $1.0125$\\ \hline
$1.00$ & $1.00$\\ \hline
\end{tabular}
\end{center}
\end{table} 
\begin{table}[htbb]
\begin{center}
\caption{$\ell=1$ state energy spectrum of the molecule ($GeV^{\alpha}$ unit)}
%$\ell=1$ state energy spectrum of the molecule (GeV unit) 
%\vspace{1mm}
%\hspace{1mm}
%\setlength{0.5}{0.
%\setlength{\tabcolsep}{0.5pt}
\renewcommand{\arraystretch}{0.7}
\centerline{
\begin{tabular}{|>{\centering\arraybackslash}m{1in}|>{\centering\arraybackslash}m{1in}|>{\centering\arraybackslash}m{1in}|>{\centering\arraybackslash}m{1in}|>{\centering\arraybackslash}m{1in}|>{\centering\arraybackslash}m{1in}|>{\centering\arraybackslash}m{1in}|}\hline
$N$ & $\alpha$ &$k_{\alpha}^*$ & $Q_1$& $\gamma_\alpha$ &$\mathcal{E}(n=1)$& $\mathcal{E}(n=2)$ \\ \hline
&$0.70$ & $1.4028884$ & $-25.3206$ &$6.5298$ & $0.9870$ &$1.2184$ \\ \cline{2-7}
&$0.75$ & $1.2795312$ & $-9.6668$ &$4.0106$ & $0.4919$ &$0.6555$ \\ \cline{2-7}
&$0.80$ & $2.2839501$ & $-4.0078$ &$2.6515$ & $0.2971$ &$0.4248$ \\ \cline{2-7}
$3$&$0.85$ & $2.1074364$ & $-3.0707$ &$2.3910$ & $0.2796$ &$0.4070$ \\ \cline{2-7}
&$0.90$ & $1.9666677$ & $-2.5078$ &$2.4630$ & $0.2141$ &$0.3100$ \\ \cline{2-7}
&$0.95$ & $2.5295886$ & $-2.4127$ &$2.5719$ & $0.2066$ &$0.2970$ \\ \cline{2-7}
&$1.00$ & $2.0652475$ & $-2.1305$ &$2.5652$ & $0.2041$ &$0.2936$ \\ \cline{1-7}
&$0.70$ & $8.2964030$ & $-10.0649$ &$3.2853$ & $0.6116$ &$0.8430$ \\ \cline{2-7}
&$0.75$ & $1.3248506$ & $-64.0555$ &$20.1808$ & $1.8151$ &$1.9788$ \\ \cline{2-7}
&$0.80$ & $4.6602700$ & $-5.3314$ & $3.1491$ & $0.3289$ & $0.4566$ \\ \cline{2-7}
$4$&$0.85$ & $7.5867908$ & $-5.6188$ &$3.3372$ & $0.3399$ &$0.4672$ \\ \cline{2-7}
&$0.90$ & $3.0443684$ & $-3.5661$ &$2.9759$ & $0.2387$ &$0.3346$ \\ \cline{2-7}
&$0.95$ & $2.8882250$ & $-3.1616$ &$2.9501$ & $0.2237$ &$0.3141$ \\ \cline{2-7}
&$1.00$ & $3.0439015$ & $-3.0988$ &$3.0494$ & $0.2258$ &$0.3153$ \\ \cline{1-7}
&$0.70$ & $8.3549641$ & $-12.4554$ &$3.7937$ & $0.6704$ &$0.9018$ \\ \cline{2-7}
&$0.75$ & $7.6900754$ & $-8.6112$ &$3.6967$ & $0.4662$ &$0.6299$ \\ \cline{2-7}
&$0.80$ & $7.0975390$ & $-6.7123$ &$3.6681$ & $0.3620$ &$0.4898$ \\ \cline{2-7}
$5$&$0.85$ & $4.4198401$ & $-5.3874$ &$3.2513$ & $0.3344$ &$0.4618$ \\ \cline{2-7}
&$0.90$ & $7.0863034$ & $-5.4862$ &$3.9065$ & $0.2833$ &$0.3793$ \\ \cline{2-7}
&$0.95$ & $3.9098650$ & $-4.1146$ &$3.4313$ & $0.2455$ &$0.3358$ \\ \cline{2-7}
&$1.00$ & $4.0396850$ & $-4.0794$ &$3.5397$ & $0.2477$ &$0.3372$ \\ \cline{1-7}
\end{tabular}
}
\end{center}
\end{table}
%\newpage 
\begin{figure}[htbp]
	\centering
		\includegraphics[width=0.75\textwidth]{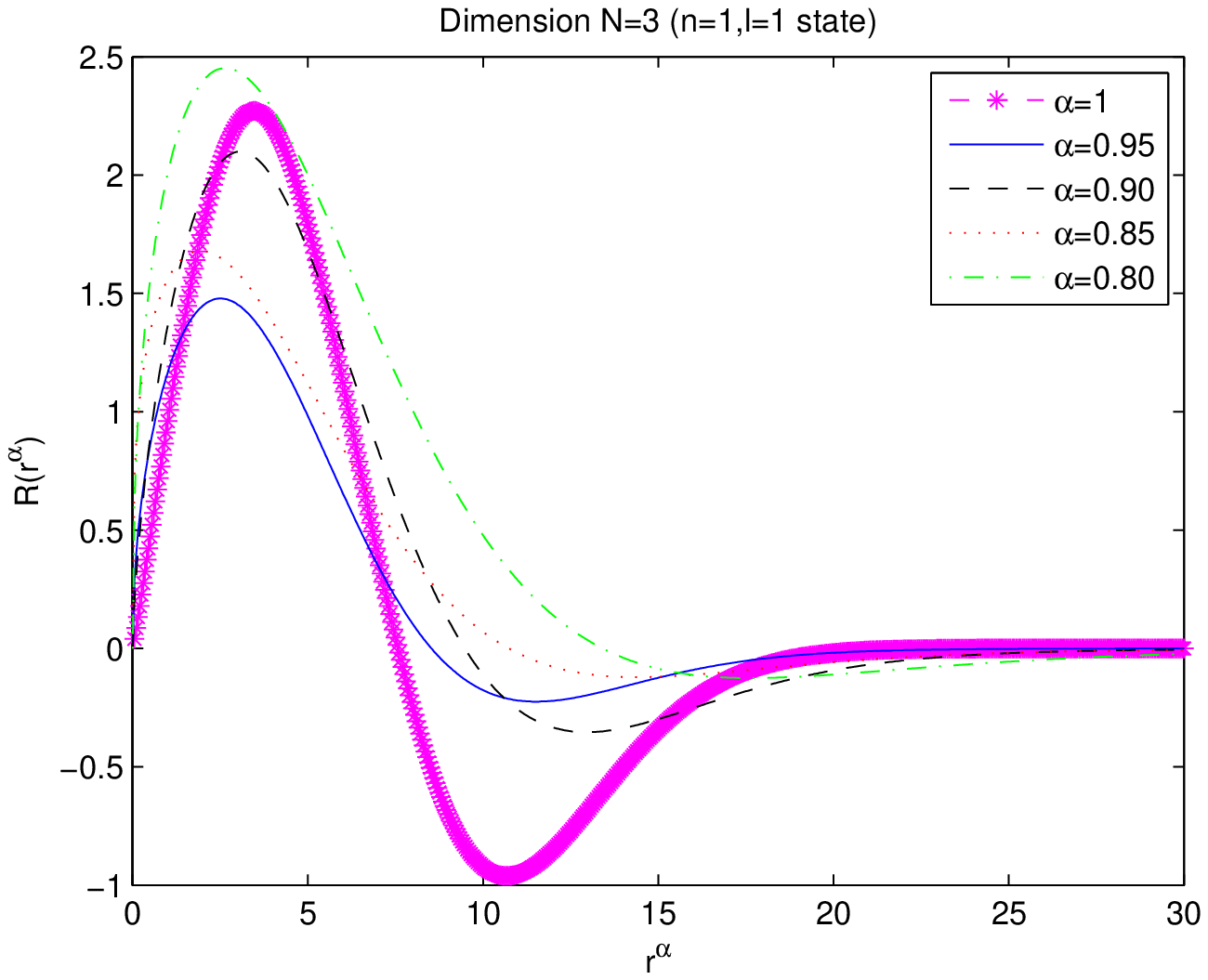}
	\caption{n=1 state eigenfunctions in N=3 dimensions for generalized fractional pseudoharmonic potential @ $\alpha=1.0,0.95,0.90,0.85,0.80$ }
	\label{fig:Fig1}
\end{figure}
\begin{figure}[htbp]
	\centering
		\includegraphics[width=0.75\textwidth]{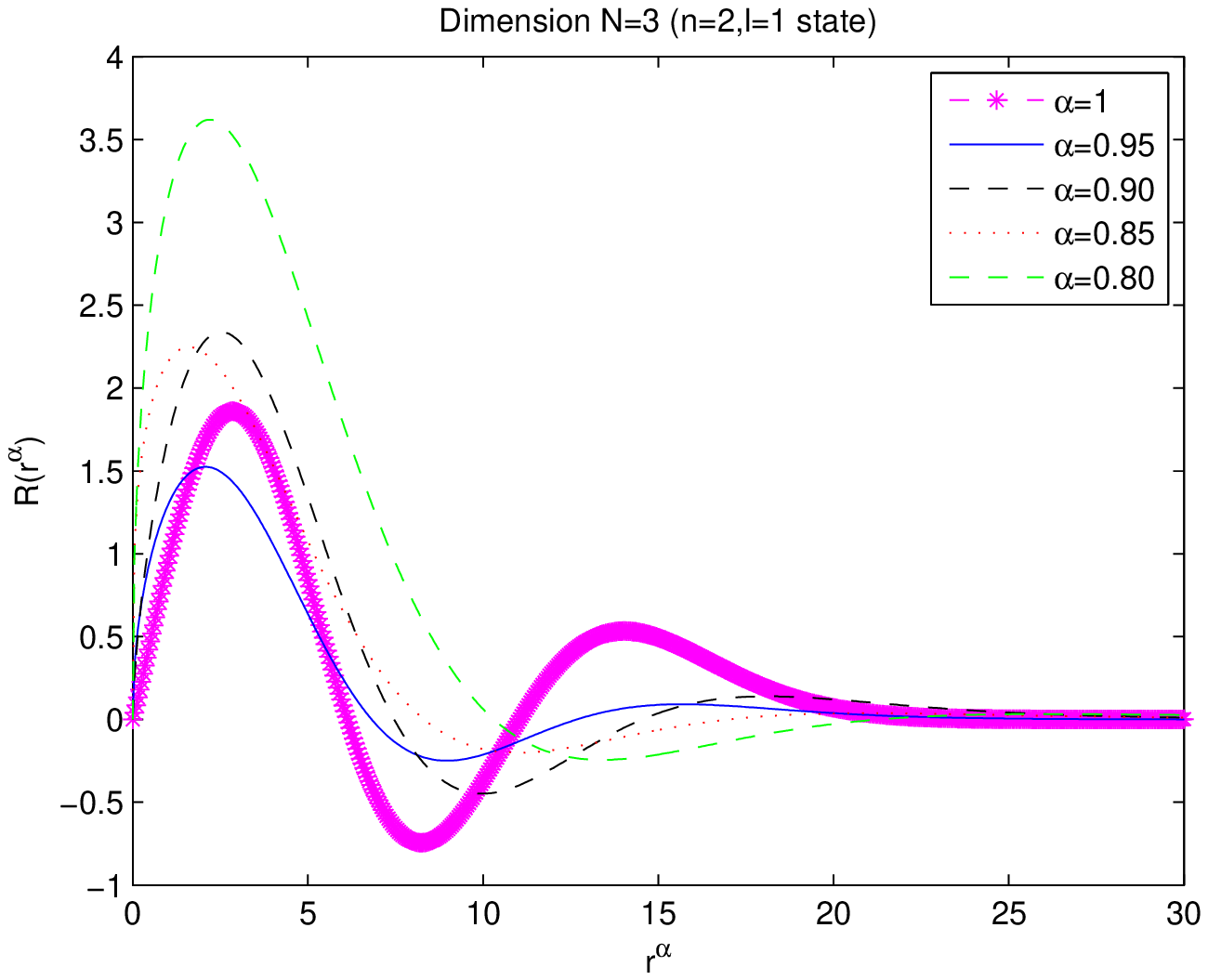}
	\caption{n=2 state eigenfunctions in N=3 dimensions for generalized fractional pseudoharmonic potential @ $\alpha=1.0,0.95,0.90,0.85, 0.80$}
	\label{fig:Fig2}
\end{figure}
\begin{figure}[htbp]
	\centering
		\includegraphics[width=0.75\textwidth]{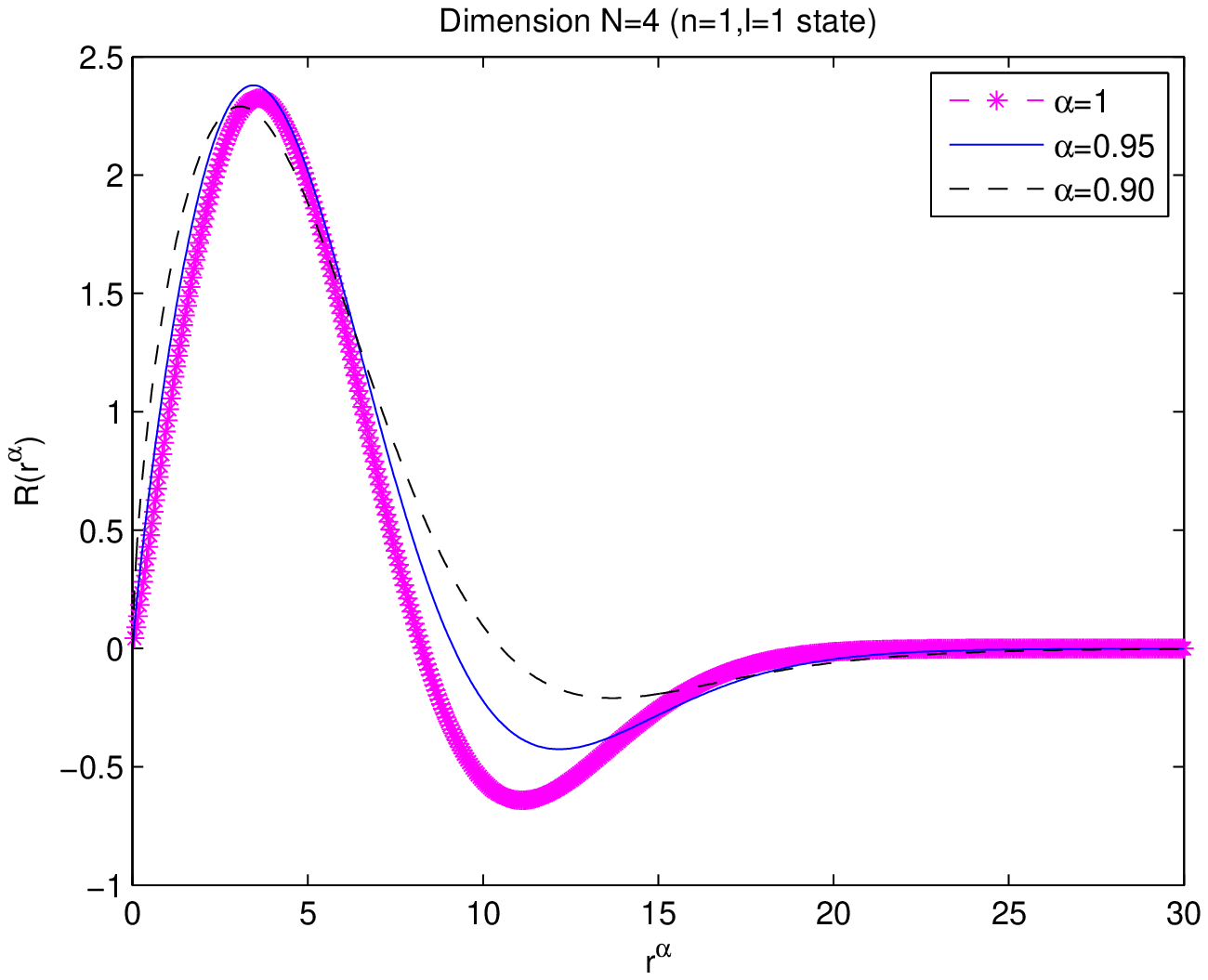}
	\caption{n=1 state eigenfunctions in N=4 dimensions for generalized fractional pseudoharmonic potential @ $\alpha=1.0,0.95,0.90$}
	\label{fig:Fig3}
\end{figure}
\begin{figure}[htbp]
	\centering
		\includegraphics[width=0.75\textwidth]{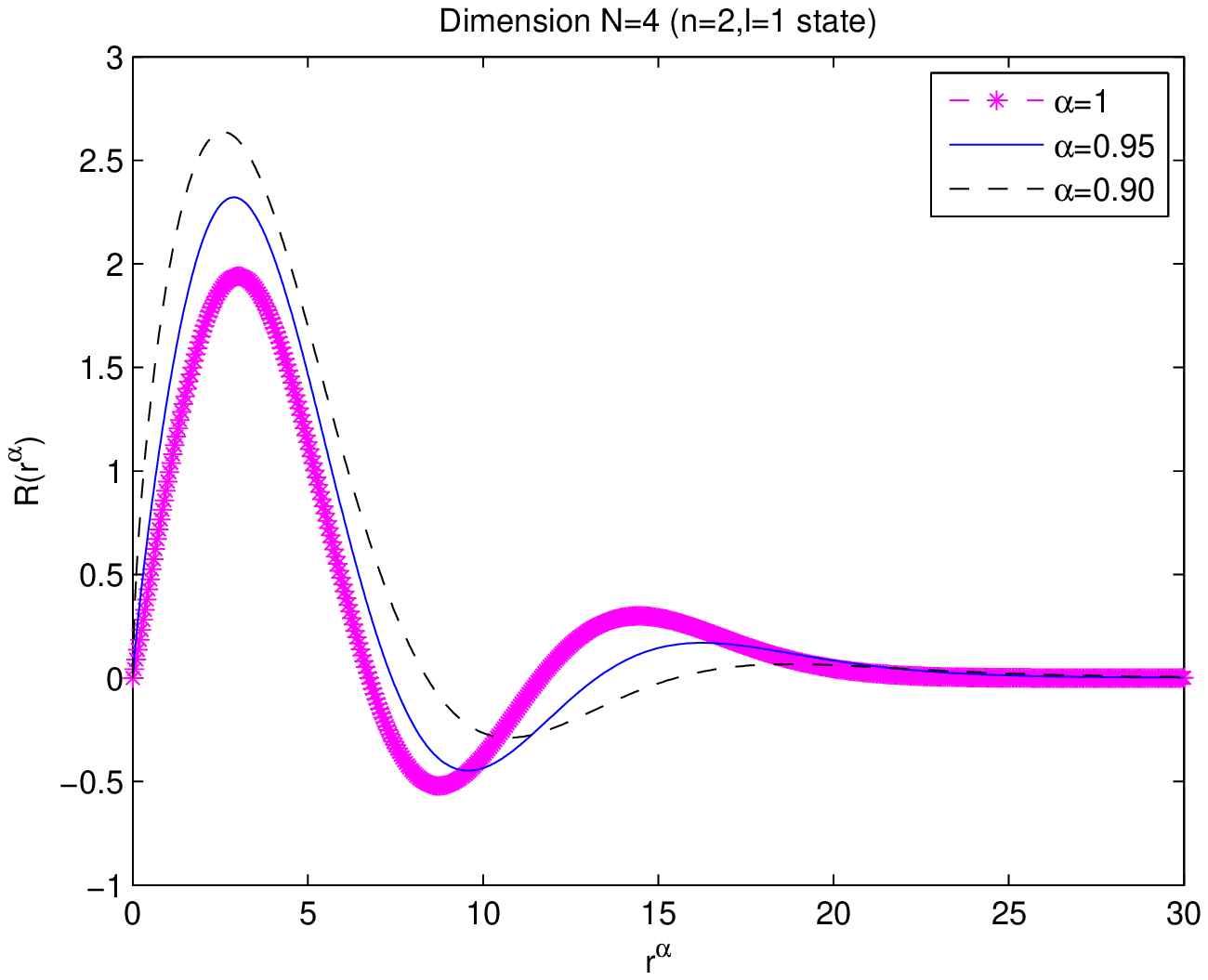}
	\caption{n=2 state eigenfunctions in N=4 dimensions for generalized fractional pseudoharmonic potential @ $\alpha=1.0,0.95,0.90$}
	\label{fig:Fig4}
\end{figure}
\begin{figure}[htbp]
	\centering
		\includegraphics[width=0.75\textwidth]{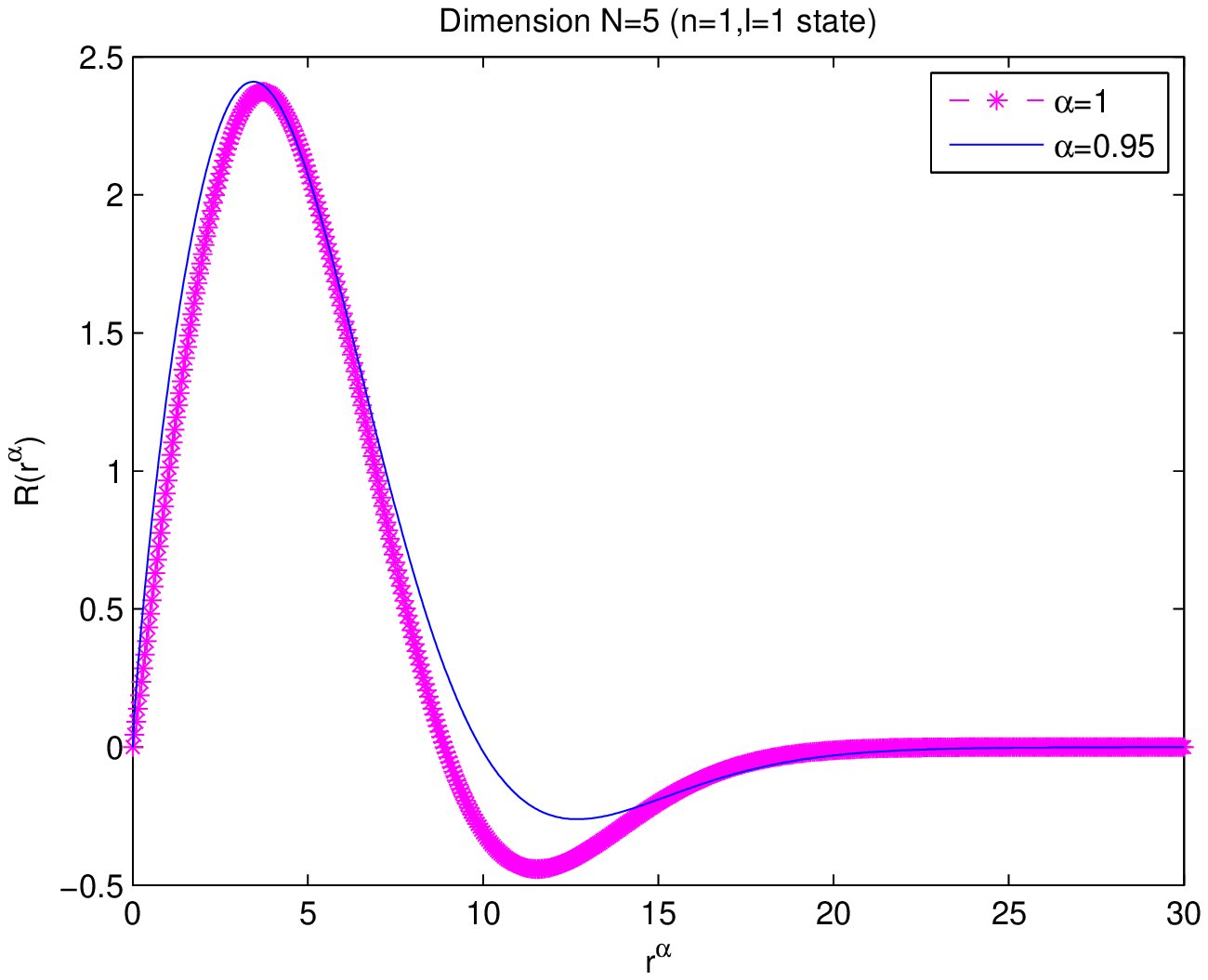}
	\caption{n=1 state eigenfunctions in N=5 dimensions for generalized fractional pseudoharmonic potential @ $\alpha=1.0,0.95$}
	\label{fig:Fig5}
\end{figure}
\begin{figure}[htbp]
	\centering
		\includegraphics[width=0.75\textwidth]{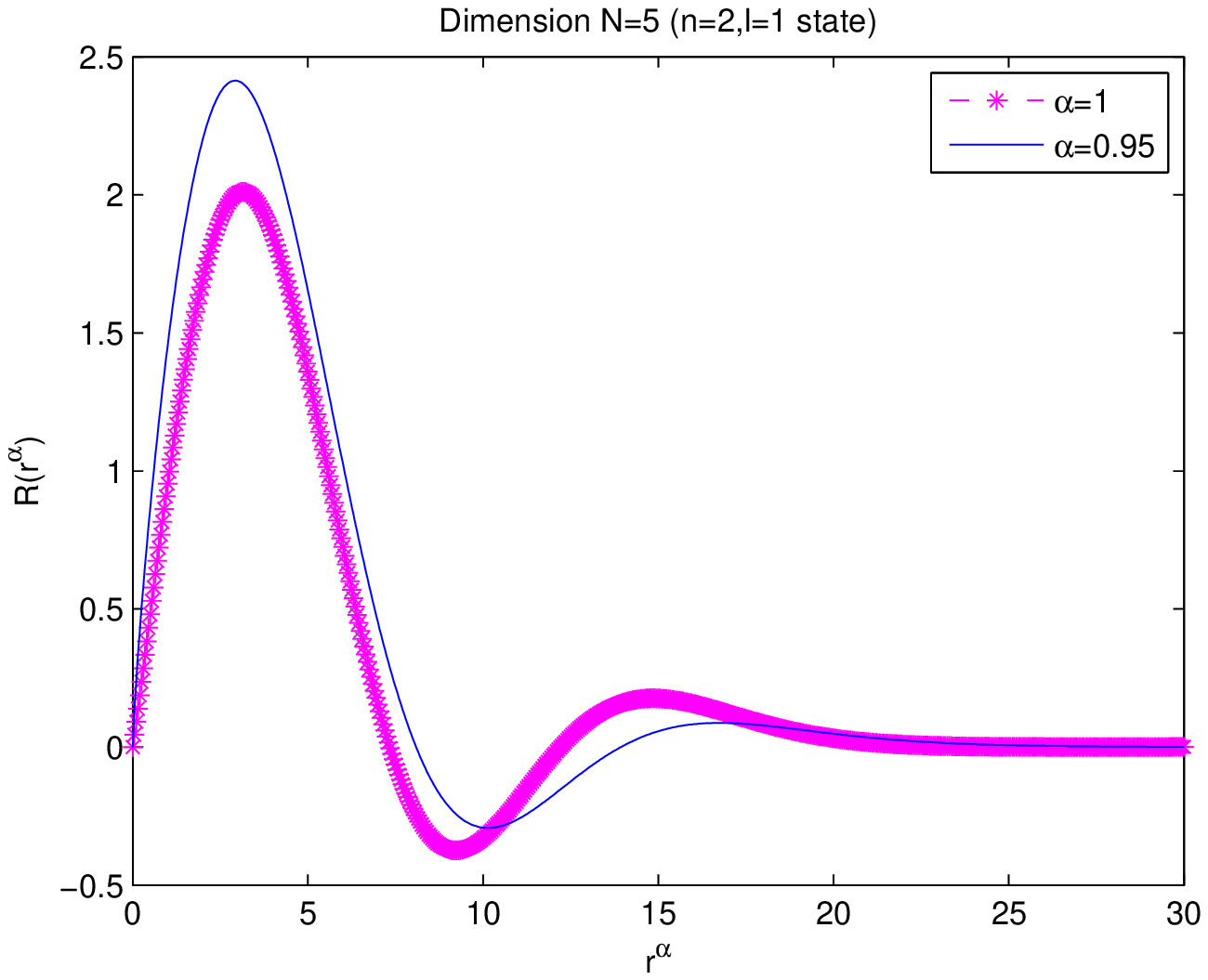}
	\caption{n=2 state eigenfunctions in N=5 dimensions for generalized fractional pseudoharmonic potential @ $\alpha=1.0,0.95$}
	\label{fig:Fig6}
\end{figure}
\begin{figure}[htbp]
	\centering
		\includegraphics[width=0.75\textwidth]{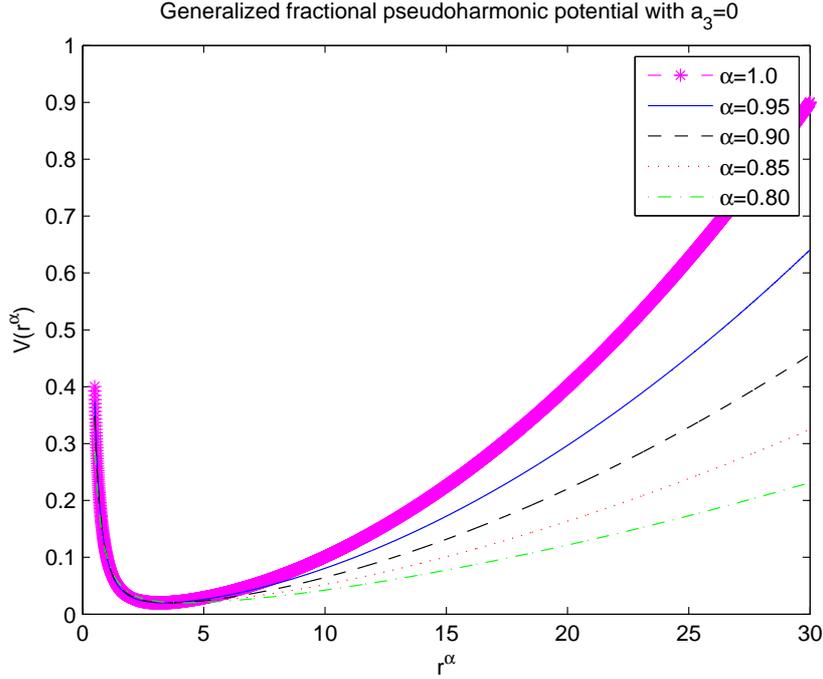}
	\caption{ Generalized fractional pseudoharmonic potential with $a_3=0$}
\label{fig:Fig7}
\end{figure}
\newpage
Above figures viz FIG.1 to FIG.6 for eigenfunctions are direct consequence of the TABLE 2. It is clear, for lower $\alpha$ far from unity, the graphs are loosing its periodicity specially for ordinary three dimensional space. There is a critical value of $\alpha$, below it the nature of the eigenfunction becomes unsuitable physically. For three dimensional space, numerical results predict the value just near about $0.8$. The critical value depends on the dimension and the potential function (FIG.7). This can be seen in FIG.3 to FIG.6  where the critical value lies more near to $\alpha=1$.    
\subsection{Review of mass spectra of Quarkonium via Cornell potential}
Though our entire model is approximated for $\alpha\approx 1.00$ but we can not resist ourselves to check the situation for $\alpha=0.5$. Taking the potential parameter $a_1=a$ and $a_2=b$ with $a_3=0$ the potential given by equation (1) becomes $V(r^\frac{1}{2})=V_c(r)=ar+\frac{b}{r}$. This is renowned Cornell potential [33], generally taken in non-relativistic quantum chromodynamics (NRQCD) for realizing the quarkonium states. Apart from the light quarks, \textit{bottom} and \textit{charm} are much heavy. The speed of the charm quark is $0.3c$ where the same for bottom is $0.1c$. Thus relativistic effects on charm and bottom are small, that is why NRQCD is enough for computing the states of quarkonium. In addition to that it is also seen that NRQCD fits much better for bottom quark due its larger mass $m_b$ compared with charm mass $m_c$.\\
Quarkonium means flavorless meson with combination of two quarks which is symbolized as $q\bar{q}$, where $q$ is quark and $\bar{q}$ is its anti quark. For heavy quarks bottom (b) and charm (c) the quarkonia are written as $b\bar{b}$ and $c\bar{c}$. The first one is bottomonium and second is charmonium. In NRQCD, study of the quarkonium states are done effectively via a static potential. The most popular potential in the list is Cornell potential $V_c(r)$. The first part i.e the linear part `$ar$' is responsible for confinement of the quarks and the second part `$\frac{b}{r}$' is usual Coulomb part which defines the e.m force between the constituent quarks in certain quarkonium.\\
Now coming back to our model, Cornell potential is obvious consequence of $\alpha=0.5$ and $a_3=0$. Since our model is approximated for $\alpha\approx 1.00$ it will not be possible to determine the exact bound state eigenfunctions as well as energy eigenvalue equation directly. We may use the present energy value equation with slight modification to investigate the situation for the mass spectra of quarkonium states. The modification means to correct the dimension or unit of the energy eigenvalue equation to incorporate the situation for $\alpha=0.5$. Once again we will use natural unit scheme here. Cornell potential with $a_1=a$ and $a_2=b$ will make the unit of $a$ as of $GeV^{2\alpha}$ since the unit of $r$ (in fractional sense) is $ GeV^{-\alpha}$ and $b$ will be unit free to make the unit of the potential $GeV^{\alpha}$. This will turn the unit of $[\mathcal{E}_{n\ell}]_{a_3=0}$ as $ GeV^{\frac{\alpha}{2}}$ since the unit of the mass is $GeV^{\alpha}$. We propose a dimensional term $\zeta$ with the unit $GeV^{\frac{\alpha}{2}}$ such that the unit of the energy value equation emerges as $GeV^{\alpha}$. The expression of $\zeta$ is unknown to us. Furthermore we assigned the value of $\zeta$ very close to unity to check the validity of the energy value equation numerically. Accepting all these we write $[\mathcal{E}_{n\ell}]_{a_3=0}\rightarrow \zeta[\mathcal{E}_{n\ell}]_{a_3=0}$ for $\alpha=0.5$. \\
The mass spectra of quarkonium state in three dimension is $M_q=2m_q+\mathcal{E}_{n\ell}$ [34]. So along with the modification proposed, we have from Eq.(3.23) 
\begin{eqnarray}
M_q=2m_q+\zeta\frac{2\tau^2}{\Gamma(3/2)}\Big(n+\frac{\gamma_{0.5}}{2}\Big)\sqrt{\frac{a}{M}}\,,
\end{eqnarray} 
where $M=\frac{m_b}{2}$ or $\frac{m_c}{2}$ according to $q$ is $b$ or $c$. The major concern factor of the above equation is $\tau=-\frac{cosec((\alpha-\delta)\pi)}{cosec(-\delta \pi)}$ with $-1<\delta<0$. At $\alpha=0.5$ it offers infinite or large value when $\delta=-0.5$. In this situation we choose $\delta=-\frac{1}{3}$ to evaluate $\tau$ for $\alpha=0.5$. Any other value of $\delta$ may give the somehow similar results but for the best results we have chosen $\delta=-\frac{1}{3}$. This can be seen in FIG.8, where the variation of $\tau$ with $\delta$ has been shown. $\delta=-\frac{1}{3}$ provides the variation much more symmetric under the range of $0<\alpha<1$.
\begin{figure}[htbp]
\centering
\includegraphics[width=0.75 \textwidth]{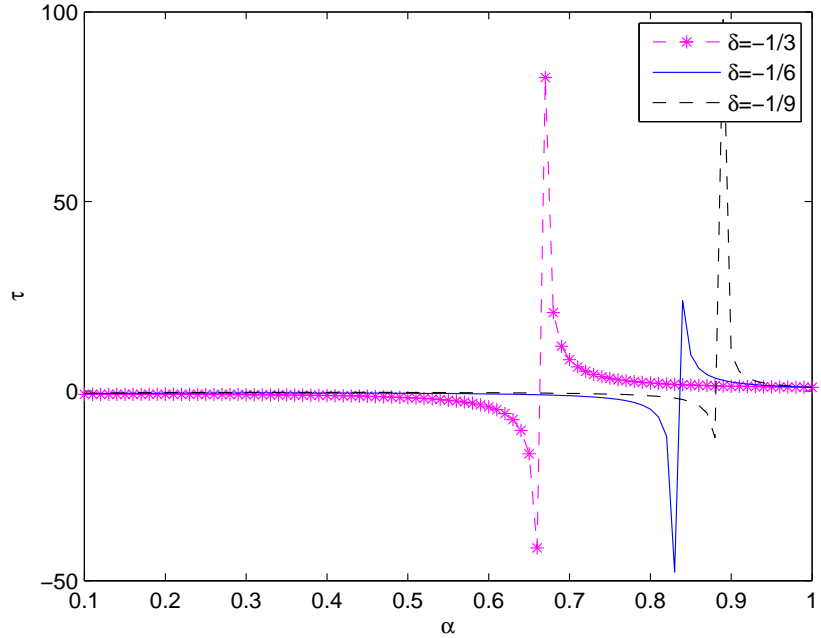}
\caption{Variation of $\tau$ with $\alpha$ when $\delta$ acts as parameter}
\label{fig:FIG8}
\end{figure}
The mass spectra of bottomonium and charmonium are calculted from Eq.(4.9) by taking standard parameter values [35]. The results are displayed in TABLE 3. The mass spectra confirms that our model is very close to the earlier experimental results in this field. More over the obtained values of mass spectra also indicates that bottomonium obeys NRQCD well than the charmonium.
\begin{table}[htbp]
\begin{center}
\caption{{\bf Quarkonium mass spectra in $GeV^{\alpha}$ unit}}
Bottomonium ($b\bar{b}$)($m_b=4.803$ $GeV^{\alpha}$, $a=0.095$ $GeV^{2\alpha}$, $b=-1.0$)\\
Charmonium  ($c\bar{c}$) ($m_c=1.480$ $GeV^{\alpha}$, $a=0.010$ $GeV^{2\alpha}$,  $b=-2.0$)\\
%\vspace{2 mm}
\renewcommand{\arraystretch}{0.8}
\begin{tabular}{|>{\centering\arraybackslash}m{1in}|>{\centering\arraybackslash}m{1in}|>{\centering\arraybackslash}m{1in}|>{\centering\arraybackslash}m{1in}|}\hline
Quarkonium & State $(n,\ell)$ & $From Eq.(4.9)$ & $Exp[36-37]$ \\ \hline
& 1S $(1,0)$ & $9.5700$ & $9.460$ \\ \cline{2-4}
$b\bar{b}$ & 1P $(1,1)$ & $9.1360$ & $9.900$ \\ \cline{2-4}
& 2S $(2,0)$ & $10.9166$ & $10.023$ \\ \cline{2-4}
& 2P $(2,1)$ & $10.4826$ & $10.260$ \\ \cline{1-4}
& 1S $(1,0)$ & $2.7053$ & $3.068$ \\ \cline{2-4}
$c\bar{c}$& 1P $(1,1)$ & $2.4289$ & $3.525$ \\ \cline{2-4}
& 2S $(2,0)$ & $3.4923$ & $3.663$ \\ \cline{2-4}
& 2P $(2,1)$ & $3.2160$ & $3.773$ \\ \cline{1-4}
\end{tabular}
\end{center}
\end{table} 
\newpage
\section{C\lowercase{onclusion}}
This present study is a sequel of our previous work which was on fractional Mie-type potential and cited in reference 27. In this paper, we have studied approximate bound state solutions of $N$ dimensional fractional Schr\"{o}dinger equation for generalised pseudoharmonic potential namely $V(r^{\alpha})=a_1r^{2\alpha}+\frac{a_2}{r^{2\alpha}}+a_3$ where $\alpha(0<\alpha<1)$ acts like a fractional parameter for the space variable $r$. We have composed the entire study by Jumarie type derivative rules with the desirability of Laplace transform . Obtained results are verified for harmonic and pseudoharmonic potentials in lower as well as in higher dimension with $\alpha=1$. We have also furnished numerical results and few eigenfunction plots for different $\alpha$ close to unity. Addition to that we have tried to obtain the mass spectra of quarkonia through the model of Cornell potential which is a special case of our potential model corresponds to $\alpha=0.5$. \\
The generalized pseudoharmonic potential for different $\alpha$ is shown in FIG 7. It is clear, as $\alpha$ goes to lower  value from unity the potential graph tends toward the $r^{\alpha}$ axis, that means the effect of potential is gradually fading. The eigenfunctions, specially for higher dimension with lower $\alpha$, are becoming more wide than the same for $\alpha=1.0$. This means the particle under the potential experiences less resistance to its motion when $\alpha$ goes with the less value than the unity. Since the motion of the particle is less affected by the potential, its position will become more uncertain and the graph will be more wider. This is what we achieved in the all figures starting from FIG 1 to FIG 7. 
%\newpage
\section*{R\lowercase{eferences}}

\end{document}